# A MANY-CORE OVERLAY FOR HIGH-PERFORMANCE EMBEDDED COMPUTING ON FPGAS

*Mário Véstias*[†], *Horácio Neto*[⋆]

[†]INESC-ID, Instituto Superior de Engenharia de Lisboa, IPL, Portugal,
[⋆]INESC-ID, Instituto Superior Técnico, Universidade de Lisboa, Portugal,
mvestias@deetc.isel.pt, hcn@inesc-id.pt

## ABSTRACT

In this work, we propose a configurable many-core overlay for high-performance embedded computing. The size of internal memory, supported operations and number of ports can be configured independently for each core of the overlay. The overlay was evaluated with matrix multiplication, LU decomposition and Fast-Fourier Transform (FFT) on a ZYNQ-7020 FPGA platform. The results show that using a system-level many-core overlay avoids complex hardware design and still provides good performance results.

## I. INTRODUCTION

Embedded systems have witnessed an exponential growth in number, diversity and processing requirements. Due to the stringent time-to-market requirements, high performance embedded systems must be flexible enough to adapt to diverse utilization cases and updates, and at the same time must have enough processing capacity to meet real-time requirements and high-performance needs. A very robust architecture in the whole design space consists of one or more configurable processing units implemented in field-programmable gate array (FPGA). FPGAs offer the benefits of determinism and reliability without the need for an application-specific integrated circuit (ASIC) and can execute run-time critical algorithms or applications that otherwise would have to be implemented in a slower and more power hungry processor.

In general, designing dedicated hardware for a specific algorithm in FPGAs requires hardware expertise. Instead of designing dedicated hardware, one approach is to consider an overlay with an underlying model of computation automatically generated and ready to be programmed and/or configured by the software developer. Constraining the hardware design space to a specific overlay may reduce the performance and increase the power consumption compared to a fully-optimized solution. However, it typically provides a good tradeoff between hardware performance, hardware portability and design time.

In this work, we consider an overlay of a many-core architecture that can be automatically generated and integrated with an embedded processor as a co-processor. To optimize the overlay, we made it configurable at two levels. At the lowest level, the designer configures the number of cores, system memory (number and size of local memories, cache and interfaces to external memory) and can also fix the topology of the interconnection network. At a higher level, the architecture can be dynamically changed without changing the lowest level architecture. In particular, the interconnection network can be changed by configuring switching circuits of the network (NoC, ring or simply point-to-point connections), the number and type of arithmetic operations of each core, number formats, including floating-point and integer (custom formats must be configured at the lowest level).

Each core has local memory, an arithmetic unit and input/output interfaces. Keeping the core simple permits to explore more parallelism and makes configuration easier. We have used three algorithms (matrix multiplication, LU decomposition and Fast-Fourier Transform) to evaluate architectures designed from the overlay.

The paper is organized as follows. Section 2 describes the state-of-the-art in many-core processing architectures for FPGAs. Section 3 describes the proposed many-core overlay. Section 4 shows the results obtained and section 5 concludes the paper.

## II. RELATED WORK

A few many-core designs on FPGA have been proposed. The MPLEM system [2] consists of Xilinx MicroBlaze soft-core processors connected with On-chip Peripheral Bus (OPB) buses. Each core has a private local memory block and all cores share an external RAM. The system synthesized in the biggest Virtex-5 FPGA (XC5VLX330T) can have 80 processors with no floating-point support. In [3] a system with 24 MicroBlaze cores interconnected with an Arteris NoC [4] was proposed. The system was implemented in a Virtex-4 FX-140 FPGA.

MARC (Many-core Approach to Reconfigurable Computing) [5] is a many-core template comprising one control processor and multiple processors for running tasks as SIMD (single instruction multiple data) units. Cores can be configured as RISC processors or synthesized as full-custom datapaths. Each core has local private memory and have access to an internal shared memory. Processors are interconnected with a network selected from a library with various topologies, including crossbar and torus. A prototype with 48 cores was implemented in a Virtex-5 FPGA. Results show that using datapath optimized processing cores the relative performance compared to a custom FPGA implementation is at most around 35% and the area efficiency goes up to 55%.

SMYLEref [9] is a many-core architecture for embedded systems prototyped in FPGA. The architecture consists of multiple clusters arranged in a two-dimensional array connected with a NoC. Each cluster has a number of scalar processors connected with a local bus. Each core has dedicated instruction and data L1 caches. A second layer of cache exists in each cluster shared by all cores. The processor core is a Geyser [10]. A prototype of the architecture in a Virtex-6 XC6VLX240T can have at most eight processors with each processor running at 10 MHz. The system was tested with an FFT and LU decomposition. For the FFT, the performance increases with the number of cores, but with 8 processors the performance efficiency drops down to around 65%. The LU performance is worse, with a performance degradation when using more than four processors.

Recently, Recore released a many-core processor subsystem for FPGAs [6]. The many-core processor connects general-purpose Xentium DSP cores and other IP blocks via a hybrid Network-on-Chip/AMBA bus. The many-core processor comes with a Software Development Environment and a functional simulator. It supports up to 4 Xentium 32-bit DSP cores, 512 kB SRAM memory tiles, connected to the NoC. The system runs at 60 MHz.

Most of these proposals rely on general-purpose embedded processors as the core unit. This increases flexibility but decreases performance and area efficiency. In approaches, like MARC, it is possible to customize the processor with a dedicated datapath that requires hardware design, but the results are still far from the peak capacity of the FPGA. In general, the communication is based on a NoC but in some approaches, e.g., MARC, it can be customized. Design space exploration is not specified in most approaches, but HeMPS, for example, uses ISS and system level simulation models to explore different platforms.

Several works have proposed dedicated many-core architectures on FPGA for high-performance computing. For example, [12] proposed recently a $12 \times 12$ systolic multiply-accumulate array for matrix multiplication reaching 169 GFLOPs. [13] implemented a linear array of 36 processing elements in a Virtex-5 FPGA for LU decomposition. The implementation with two DMA channels achieved a sustained performance of 8.5 GFLOPs corresponding to an efficiency of 89%. All these are dedicated many-core architectures for particular algorithms.

Recently, the viability of a GPU-like overlay for FPGA was analyzed [15]. However, whether GPU-like programming models and architectures are a good way to design many-cores on FPGA is yet to be checked.

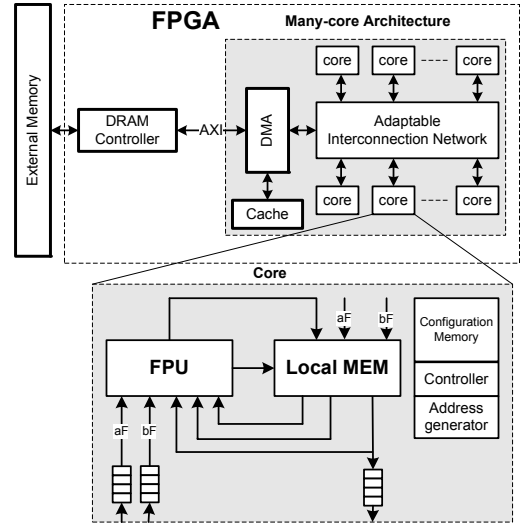

**Fig. 1**. Many-core architecture

Our objective is to design many-core architectures to work as co-processors of a general-purpose processor for high performance embedded systems. The co-processors are used for data intensive processing. To design such architectures we propose a many-core overlay that can be configured to support the execution of different algorithms in isolation or in the same many-core architecture. The core elements are based on simple processing units with reduced control, small local memories and arithmetic units. Each core unit can be individually configured in terms of local memory size and number and type of arithmetic operations. This permits to improve performance and area efficiency when compared to many-core architectures based on general-purpose embedded processors. We consider a customizable interconnection network that can be a bus, a crossbar, a NoC or a ring, and that can use point-to-point connections and/or a mix of these topologies.

### III. MANY-CORE ARCHITECTURE

The proposed many-core overlay is configurable to any number of cores (see Figure 1).

Each core has an arithmetic unit and a local data memory. The arithmetic unit can be statically or dynamically configured to execute a set of functions: add/sub, multiplier, fused multiply-add, reciprocal, square root and inverse square-root [8]. Each core can be configured with a different combination of operations. The local memory is implemented with dual-port block RAMs that are used to store temporary variables, coefficients to implement some of the arithmetic operators, constants, and output data.

The many-core has access to external memory through a DMA that is configured by the embedded processor. The DMA is responsible for sending/receiving data to/from memory and for forwarding this data to the network. In order to improve the bandwidth when requesting elements stored non-sequentially in memory, the DMA has a cache

72

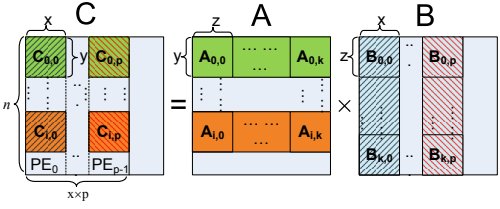

**Fig. 2**. Block multiplication algorithm.

to buffer bursts of data and thus enable faster access. Each time non-sequential data is requested from memory, a burst of sequentially-stored elements is fetched (cacheline size). The first element of the burst is the data requested. This data is immediately forwarded to the processors. The other elements are stored in cache. This is useful, for example, in algorithms working with vectors. The cache can be configured in terms of size and size of the cacheline. Each entry of the cacheline is a word whose size depends on the data width of the arithmetic unit.

The cores are interconnected with an adaptable network that can be configured statically as a bus, a crossbar, a NoC, a ring, point-to-point connections or a mix of these topologies. Alternativelly, a generic interconnection network can be used with configurable switches that can be adapted to communication requirements without architectural changes. The cores are connected to the communication network through two input and one output buffers.

## IV. RESULTS

To evaluate the architecture, we have considered parallel algorithms for dense matrix multiplication, LU decomposition and FFT all with data represented in single precision floating-point. In all cases, the design space was explored using SystemC models of the architecture and the algorithms [16] looking for the best many-core. Both architectures were implemented and tested in the reconfigurable area of a XCZ7020 SoC FPGA.

### IV-A. Configuration of the Many-Core Overlay for Matrix Multiplication

Matrix multiplication $C = A \times B$ is implemented as a parallel block matrix algorithm that partitions $C$ matrix into smaller sub-matrices (blocks) and works with these blocks. All matrices are square and have the same size ($n \times n$).

As illustrated in figure 2, the $C$ matrix is divided in blocks with size $n \times xp$. Each of these blocks is calculated by p cores simultaneously. Each core is responsible for a sub block with size $n \times x$ which in turn is divided in smaller blocks with dimension $y \times x$. The size of these smaller blocks, $C_{ij}$, depends on the local memory size. To generate a block $C_{ij}$ the processor multiplies a block $y \times n$ from matrix A with a block $n \times x$ from matrix B. The multiplication is implemented as a sequence of k partial block multiplications,

$$C_{ij} = \sum_{k=1}^{k_0} A_{ik} \times B_{kj} \quad (1)$$

Each partial block is the multiplication of a $y \times z$ sub block $A_{ik}$ with a $z \times x$ sub block $B_{kj}$, resulting in a partial sub block $C_{ij_k}$ of size $y \times x$. The final $C_{ij}$ result is obtained after accumulating the $k$ partial block results.

The partial block multiplications are implemented as follows. First, each PE receives and stores its $B_{qj}$ elements. Then, $A_{iq}$ elements are broadcasted to all processors. As the $A_{iq}$ elements arrive, they are multiplied by all $B_{qj}$ elements stored in local memory. The partial results of each block $C_{ij}$ are also stored in local memory. In the final iteration, the elements of the result block $C_{ij}$ are sent to the external memory. As referred, the local memory in each processor must store the blocks of B (size $z \times x$) and C (size $x \times y$) under processing.

At the algorithmic level, $x$, $y$ and $z$ are variables and thus different performance results are obtained by changing these values. To optimize the final solution, we have considered the theoretical results in [11] to determine these values. According to the referenced theoretical results, the number of communications with the external memory does not depend on the dimension $z$ of the sub blocks. Therefore, $z$ can be simply made equal to 1 in order to reduce the local memory required. The local memory necessary to store the sub blocks of B (size $1 \times x$) is doubled in order to enable the processor to store a new $B$ sub-block while still performing the computations with the former $B$ sub-block.

Also according to this reference, the dimensions of the sub blocks $C_{ij}$ that minimize the number of communications, as a function of the available local memory $L$, are

$$x = \frac{L}{2 + \sqrt{p\,L}} \quad y = \sqrt{p\,L} \quad (2)$$

At the architectural level, matrix multiplication requires multiply and add operations. So, the arithmetic units of all cores are configured as fused multiply-add. The cores are connected in a linear array, that is, each core is connected to two neighbors with point-to-point links.

We have configured the overlay with 16 and 32 cores, all with the same local memory size and a DMA cache with support for up to 16 cachelines. We have configured the overlay as a one-dimensional array of the cores connected to external memory through the DMA.

For both architectures an initial study was done to find the relation between cache line and local memory sizes (see table I).

With 16 cores, local memories of 2 KBytes (per each core) and a cache with a cache line of size 16 are enough to implement the architecture with the best performance. In the case of 32 cores, the cache line must be doubled to achieve the best performance. With smaller local memories, than those indicated in the table, the external memory access requirements increase such that the communications cycles become higher than the processing cycles and therefore the final performance is always worst independently of the size of the cacheline.

Table I. Relation between cache line size and local memory size to guarantee best performance

| 16 cores | | | | |
|---|---|---|---|---|
| Local Memory | Cacheline | y | Cache size | Total Memory |
| 32 KBytes | 1 | 256 | 1KByte | 513 KBytes |
| 16 KBytes | 2 | 256 | 2 KByte | 258 KBytes |
| 8 KBytes | 4 | 256 | 4 KByte | 132 KBytes |
| 4 KByte | 8 | 128 | 4 KByte | 68 KBytes |
| 2 KBytes | 16 | 128 | 8 KByte | 40 KBytes |
| 32 cores | | | | |
| Local Memory | Cacheline | y | Cache Size | Total Memory |
| 16 KBytes | 2 | 256 | 2 KBytes | 514 KBytes |
| 8 KBytes | 8 | 256 | 8 KBytes | 264 KBytes |
| 4 KBytes | 16 | 256 | 16 KBytes | 144 KBytes |

Assuming an architecture with 32 KBytes of local memory for the 16-core and 16 KBytes for the 32-core architecture, we have determined the utilization of resources and the number of execution cycles (see table II). Both architectures achieve high performance efficiencies (peak performance/measured performance), 86% and 84%, respectively. The 16-core achieves 7 GFLOPs and the 32-core achieves 13.4 GFLOPs.

Table II. Results for matrix multiplication

| | Core | Arch. 16-cores | Arch. 32-cores |
|---|---|---|---|
| LUTs | 1,364 | 24,390 | 46,576 |
| DSPs | 4 | 71 | 135 |
| BRAMs | 8/4 | 140 | 140 |
| Freq. (MHz) | 250 | 250 | 250 |
| Cycles | — | 77,772,668 | 39,796,887 |
| Time (s) | — | 0.31 | 0.16 |
| GFLOPs | 0.5 | 7 | 13.5 |
| Efficiency | — | 86% | 84% |

Compared to previous implementations, ours has about half of the performance of the dedicated architecture for matrix multiplication in [14], but consumes only about 25% of the resources. Doubling the number of cores of our architecture would provide an architecture with the same performance. In terms of efficiency, our architecture is better. We also have higher efficiencies compared to the dedicated many-core proposed in [12].

### IV-B. Configuration of the Many-Core Overlay for LU decomposition

LU Decomposition factors a square matrix, of size $n \times n$, into an upper triangular matrix, U, and a lower triangular matrix, L. Lower/upper triangular decomposition is performed by a sequence of Gaussian eliminations to form A=LU. A pseudocode for the column-oriented LU decomposition is given in Listing 1.

Listing 1. LU Decomposition
```
for k = 1 to n-1
  rec_a = 1/a(k,k);
  for s = k+1 to n
    l(s,k)=a(s,k) * rec_a;
  for j = k+1 to n
    for i = k+1 to n
      a(i,j) = a(i,j) - l(i,k) * a(k,j);
```

To parallelize the algorithm, we proceeded as follows. One core receives the first column of A, determines the first column of $L$ and stores it in its local memory. Next, the same processor receives the next column of $A$ and calculates the second column of $A'$ and so on, until it obtains the complete $A'$. In this process, the first element of each column of $A'$ is sent back to memory since these are already the final values of matrix $U$. As the core calculates each column of $A'$, it sends it to the next core that will calculate $A''$ following the same process. It calculates the second column of $L$, stores it in its local memory and calculates the remaining columns of $A''$ and sends them to the next core to calculate $A'''$. If $n$ is higher than the number of cores, then the last core must send its results back to global memory to be read (again) by the first core.

At the architectural level, the LU algorithm requires multiply-add and reciprocal operations. So, the arithmetic unit of all cores was configured with a fused multiply-add and a reciprocal unit. The cores are connected again as a linear array and the results are written back to memory through a bus. The DMA does not need a cache since the access is always by columns (assuming the matrix is stored by columns). Each core must store the values of matrix $L$ to be used in the calculations of new columns. Thus, in the worst case each core needs a local memory with size equal to $n - pos$, where $pos$ is its position in the chain of cores.

To simplify the implementation, we configured the local memories of all cores with the same size, 16 KBytes, enough to store at most $4K \times 32$-bit elements. We have determined the resources consumed by a single core and by the whole 16- and 32-core architectures (see table III).

Table III. Resource utilization for LU decomposition

| | Core | Arch. 16-cores | Arch. 32-cores |
|---|---|---|---|
| LUTs | 1,503 | 26,614 | 46,576 |
| DSPs | 4 | 71 | 135 |
| BRAMs | 2 | 47 | 63 |
| Freq. (MHz) | 250 | 250 | 250 |

Table IV shows the performance results for both 16- and 32-core architectures and for different matrix sizes.

The performance efficiency decreases slightly with the number of cores. The increase in the number of cycles with the size of the matrix is in accordance with the complexity of the algorithm ($n^3$). In all cases, using a second DMA channel, the efficiencies would double since the communications (and also the total execution times) would reduce 50%.

The computation time is in the order of several miliseconds, which is much better than the LU implementation in the (general-purpose) many-core architecture of [9] (computation time in the order of seconds). This is mainly due to their many-core approach being based on a complex and slower general-purpose core which for these type of applications can be quite inefficient. Our configurable architecture is also competitive with dedicated approaches. For example, a 36-core implementation in a Virtex-5, with 2 DMA controllers, of the dedicated many-core in [13], achieves 8.5 GFLOPs at a 89% performance efficiency. Our 32-core architecture, also with two DMA channels, can achieve 15 GFLOPs with a 92% performance efficicency.

Table IV. Performance results for LU decomposition

| 16 cores | | | |
|---|---|---|---|
| Size | Cycles | # operations | Efficiency |
| $128 \times 128$ | 104,017 | 699,008 | 42 % |
| $256 \times 256$ | 765,216 | 5,559,680 | 45 % |
| $512 \times 512$ | 5,853,972 | 44,739,072 | 48 % |
| 32 cores | | | |
| Size | Cycles | # operations | Efficiency |
| $128 \times 128$ | 61,164 | 699,008 | 36 % |
| $256 \times 256$ | 416,824 | 5,559,680 | 42 % |
| $512 \times 512$ | 3,061,743 | 44,739,072 | 46 % |

Table V. Performance results for FFT

| # points | Cycles | | | |
|---|---|---|---|---|
| | 4 cores | 8 cores | 16 cores | 32 cores |
| 16 | 83 | 76 | 76 | 76 |
| 32 | 179 | 144 | 144 | 144 |
| 64 | 407 | 311 | 276 | 276 |
| 128 | 899 | 667 | 536 | 536 |
| 256 | 1.991 | 1.375 | 1.052 | 1.052 |
| 512 | 4.355 | 2.819 | 2.080 | 2.080 |
| 1K | 9.479 | 6.407 | 4.871 | 4.132 |
| 2K | 20.483 | 13.579 | 10.507 | 8.232 |

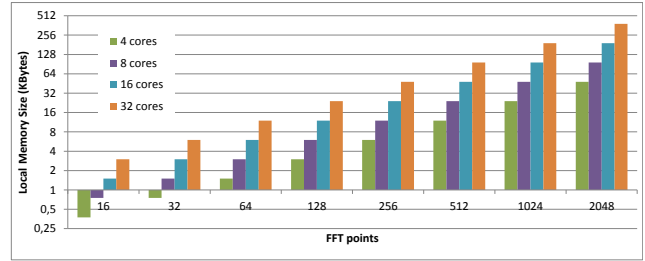

**Fig. 3**. Memory Size required for different number of cores and different number of FFT points.

### IV-C. Fast-Fourier Transform

Fast Fourier Transform is one of the most frequently used kernels in a wide variety of image and signal processing applications. Several FFT algorithms have been proposed and developed. Radix-X Cooley-Tukey algorithm is one of the most popular algorithms for hardware implementation. Starting with the basic equation of an N-point DFT (see equation 3)

$$X_p = \sum_{n=0}^{N-1} x_n \, e^{-j \frac{2\pi}{N} np} \qquad (3)$$

we partition the DFT into odd and even-indexed terms

$$\begin{aligned} X_p &= \sum_{n=0}^{\frac{N}{2}-1} x_{2n} \, e^{-j \frac{2\pi}{N} 2np} \\ &+ e^{-j \frac{2\pi}{N} p} \sum_{n=0}^{\frac{N}{2}-1} x_{2n+1} \, e^{-j \frac{2\pi}{N} 2np} \\ &= A_p + W^p B_p \end{aligned} \qquad (4)$$

$A_p$ and $B_p$ are themselves DFTs of length N/2. Hence, the FFT algorithm performs two independent N/2-point FFT and combine the results using N multiply-add operations. This gives rise to the well known complexity of the FFT algorithm, $O(nlog_2n)$.

Evaluating equation 4 at frequencies $N + \frac{p}{2}$, we obtain $X_{p+\frac{N}{2}} = A_p - W^p B_p$. Both $X_p$ and $X_{p+\frac{N}{2}}$ are calculated with a butterfly structure.

The parallel FFT algorithm implemented uses two cores to calculate each stage of the FFT, one is responsible for the real part and the other for the imaginary part. The first two cores receive the point values from memory calculates the first stage and sends the results to the second pair of cores and so on. The final results are sent to the memory. In case the number of pairs of cores are less than the number of stages, some pairs of cores have to calculates more than one stage of the FFT. For example, with N = 16, the FFT has four stages, each with eight butterflies, and therefore eight cores are used in this case.

The FFT algorithm requires only multiply and add operations and, so, the arithmetic unit of all cores is configured with a fused multiply-add, like in the architecture for matrix multiplication. All cores were configured with the same local memory. The interconnection network is more tricky. Each core is connected to two neighbors with a point-to-point connection and the results are written back to memory through a bus. The DMA does not need a cache since the access is always by columns (assuming the FFT points are stored by columns). However, we have considered two DMA channels to retrieve/save both real and complex values in the external memory.

Table V shows the performance of the architecture for different number of cores and for different FFT sizes. The results are consistent with the complexity of the FFT.

Since the algorithm considered to parallelize the FFT assumes all coefficients are stored in local memory of each processor, the supported number of FFT points will depend on the available memory. We have determined the relation between the number of FFT points and local memory size (see Figure 3).

As expected, the available memory of the target device determines the maximum number of FFT points according to the behavior illustrated in the figure. Given the limit of available internal memory, we may have to reduce the number of cores to support the execution of the algorithm, with consequent reduction of performance. After reaching the limit of available memory, higher number of FFT-points

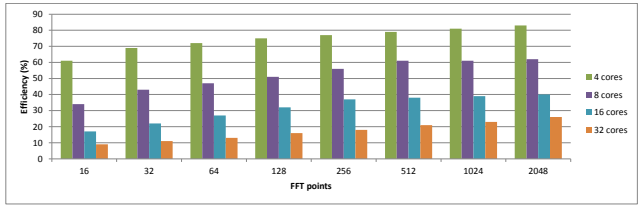

**Fig. 4**. Performance efficiency for different number of cores and different number of FFT points.

can be supported only using external memory. In this case, the communication to external memory will degrade the performance of the algorithm and possibly would be better to reduce the number of cores (this was not yet quantified)

A second aspect of the FFT implementation is the performance efficiency of the architectures (see Figure 4).

As we can see, the performance efficiency decreases with the number of cores and increases with the number of FFT points. This is also true for matrix multiplication and LU decomposition. Therefore, if we have to run some of these algorithms within a single application it is better to run them in parallel with less number of cores allocated for each algorithm than running them with all cores allocated to each algorithm serially.

We have also considered the configuration of the overlay for all three applications at once. In this case, the arithmetic units were configured to support fused multiply-operations and reciprocal. The reciprocal operation is configured dynamically by loading the coefficients of the piecewise polynomial approximation used to calculate it [8]. The interconnection network is configured statically with configurable switches that are dynamically adapted to the communication requirements of the applications. The performance results are close to those obtained with independent architectures.

## V. CONCLUSION

A configurable many-core overlay for high-performance embedded computing was proposed. Cores and interconnection topology can be configured at two levels to optimize the architecture for particular algorithms.

Previous proposals of many-core architectures for embedded systems are based on general-purpose embedded processors. Compared to our many-core, these systems in general have a better support to run control intensive kernels or threads but are less efficient for data intensive applications in terms of performance and area. This is because our cores are simpler and application optimized, and can also support higher operating frequencies.

We have evaluated the architecture for matrix multiplication, LU decomposition and FFT. The results show that the architectures generated from the many-core overlay achieves performances close to those of state-of-the-art dedicated circuits and performance efficiencies near 90% without requiring hardware expertise to design the many-core architecture.

### ACKNOWLEDGMENT

This work was supported by national funds through FCT, Fundação para a Ciência e Tecnologia, under projects PEst-OE/EEI/LA0021/2013 and PTDC/EEA-ELC/122098/2010.